# FOOTBALL AND EXTERNALITIES: USING MATHEMATICAL MODELLING TO PREDICT THE CHANGING FORTUNES OF NEWCASTLE UNITED

AJUNI SINGH | PRASHANT YADAV | VISHIST SRIVASTAVA

INDIAN INSTITUTE OF MANAGEMENT, INDORE

# ABSTRACT

The Public Investment Fund (PIF), is Saudi Arabia's sovereign wealth fund. It is one of the world's largest sovereign wealth funds, with an estimated net capital of $382 billion. It was established to invest funds on behalf of the Government of Saudi Arabia. Saudi Arabia is aiming to transfer the PIF from a mere local authority to the world's largest sovereign fund. Thus, PIF is working to manage $400 billion worth of assets by 2020.

It was with this Public Investment Fund that Saudi Arabia decided to buy out the football club- Newcastle United FC- a mid-table club of the premier league. It was earlier owned by Mike Ashley – the owner of SportsDirect.com. Mike Ashley was willing to sell the club and the Saudi's were termed as the favourites for the takeover. But, for some reasons, this did not materialise.

In this paper, we aim to forecast the investment levels and the subsequent improve in the league position of Newcastle United FC using the model of another premier league club- Manchester City as the base. We employ the DiD approach of logistical regression through Python.

Further, we also discuss the reasons which led to the collapse of this deal and the externalities at play. One main reason here was the Coronavirus outbreak which has been discussed below

## About Newcastle United F.C.

The Newcastle United Football Club is based around Tyne, Tyne and Wear in Newcastle with the British professional football team plays the highest flight of the Premier League for English football. A combination created The Newcastle lower east side and Newcastle west side in 1892. St. James' Park is the arena where the team plays their matches from home. As per the Taylor study, in the mid-1990s the ground size had been increased to 52.305, all top clubs had to be an all-seater stadium.

As of July 2020, the Club has been in top flight for 88 years, and hasn't been out of English football's second tier at joining the soccer league in 1893. For all but three years of its history the Club has been a member of the top division. Throughout the competition Newcastle was awarded four titles, six F.A. Cups and a community shield, including the 1969 inter-city fairs cup and the 2006 UEFA Intertoto Cup, a premier league club 's 9th-highest award number won. Newcastle was relegated in 2009, and again in 2016. In 2010 and 2017 the Club was promoted to top division, winning both championships the very same season.

Newcastle has a long-standing rivalry mostly with Sunderland Team, that has clashed with Tyne-Wear battle since 1898. The club kit 's regular colours, black and white, striped shirts and black pants. Their crest has attributes of city-wrap, with two brown seahorses. Before each home game and inspiring songs like "Blaydon Races" are sung the staff enter the "Town Hero" group.

The Club has belonged to Mike Ashley, successor to Sir John Hall since 2007. The Club is the 17th largest selling club in the world, producing 169.3 million Euros in 2015 as far as its annual turnover is concerned. Newcastle was the fifth biggest football club in the world in 1999 and the second-largest one in England after Manchester United.

## Summary & Timeline of the Transfer

A consortium headed by the Saudi Arabia Public Wealth Fund was promised to purchase Newcastle United in April 2020. Sales have also triggered doubts and protests, such as arguments that take the human rights history of Athletes in Canada into account, and alleged theft in the region's sports broadcasting.

In May 2020, two conservative MPs called for the government to investigate sales aspects; Karl McCartny asked for the sales stop; Giles Watling called on the Internet Technology, the Culture, Media and Leisure Department of Saudi Arabia to send an evidence conference.

In June 2020 Richard Masters, who appeared in front of the Department of Digital, Educational, Media and Leisure, suggested the purchase of Newcastle Unite. However, the MPs warns about the "humiliation" of having the Saudi group to take over given its record on piracy and human rights. The Guardian reported in July 2020, that Newcastle United had further complicated its decision to ban the broadcasting of IN Sports in the region.

Saudi Arabia declared on July 30 2020 its withdrawal from the Newcastle Agreement, after numerous media reports which emphasize realms as the key violator of human rights. In order to encourage the pirate push, the WTO decided to use Piracy-Pay-Service-BoutQ. The group said in the retreat declaration that "we decided to remove our intention to buy the Newcastle United Football Club with deep regard for Newcastle and the integrity of its club."

All the major reasons have been discussed in detail in the section "**What all factors made the Saudi PIF drop out of the Newcastle deal**".

## Breaking down the problem

The final rankings in the Premier League are determined on the basis of the points tally of each team. The number of goals which a team scores and the number of goals conceded are also recorded for each team. Their goal difference is then used to break down ties. To predict the final ranking of Newcastle United, we needed to predict the ranks of all the teams in the league. This reduced our problem to predicting the results of individual matches.

To simplify our model, we assumed that the outcomes of any two matches in the league are independent of each other, that is, the result of a match X has no influence on the result of match Y.

A match between two teams can have three possible outcomes:

1. The Home team wins: The number of goals scored by the team playing on their home ground is greater than the number of goals scored by the opponent team.
2. The Away team wins: The number of goals scored by the Away team is greater than the number of goals scored by the Home team.
3. Draw/Tie: Both the teams score the same number of goals.

We could have used a randomiser to get the match results but we realised that it might not be an accurate representation of the real-world scenario. In reality, a top tier team always has a higher chance of winning than a low tier team. So, we decided that we needed some parameters to measure the strength of a team.

## Obtaining data and choosing parameters

To measure a team's performance, we decided to use the past data available on the internet. We used the datasets available at http://www.football-data.co.uk/data. The datasets were divided into CSV files containing the results of Premier League matches from 1992 to 2020.

After doing some research and looking at other football score prediction projects, we arrived at the following conclusions:

1. Features like the number of fouls, red/yellow cards and the corners had weak correlation with the points scored and hence, the strength of the team.
2. Goal difference had the highest correlation with Team strength, this is because it basically tells us about the balance between a team's attack and defence.
3. One shocking finding was that the number of shots taken is inversely correlated with the points tally. This means that the more the number of shots a team takes, the lesser points it will have. Even though this might sound strange, this might be explained by the fact that whenever a team takes a shot which does not convert into a goal, the possession goes back to the opponent team, giving them a chance to try a counter-attack.

Therefore, we decided to base our parameter around the goal difference of both the teams as a way to quantify their attacking and defensive strengths.

At the end of the day, the points a team achieves are directly dependent upon the number of goals scored by both the teams. Hence, we decided to use the probability distribution of the number of goals scored. The best way to do this was via a Poisson distribution.

The Poisson distribution estimates the likelihood of a given number of events happening in a fixed period of time if these events occur with an identified constant rate as well as are independent of the time since the last event.

$$P(X = x) = \frac{\lambda^x e^{-\lambda}}{x!}$$

To depict why Poisson distribution works for our project, we came up with the following example. We considered each goal scored as an independent event. Then, within a match of 90 minutes, each scoring event occurs any number of times independently. We tried to find out what are the chances that a match between Arsenal and Leicester City concludes with the score line 2-1.

```
Let's say that following figures are obtained:
Average number of goals scored by Arsenal = 1.842
Average number of goals scored by Leicester City = 1.500
Then,
A) Probability that Arsenal scores 2 goals = P(X=2,λ=1.842) = 27.8%
B) Probability that Leicester scores 1 goal = P(X=1,λ=1.5) = 33.4%

Hence, probability of the scoreline 2-1 = (A)*(B) = 9.2%
```

The only problem left was to figure out the constant rate (λ).

It can be instinctively seen that this parameter should account for the performance of the team. The better team has a higher chance of scoring goals. The number of goals scored also depends on the defensive strength of the opponent team. Lastly, the 'Home Advantage' also plays a major role in influencing the performance of a team, this means that there should be separate parameters for the teams involved in a match based on the venue.

Thus, we defined the parameter λ as the Average number of goals scored by a team on a particular venue, which we computed using the available past data.

## Building the model

To derive a separate constant rate constant rate (λ) for both home and away matches, we decided to use the following parameters:

```
For each team, we can compute:
a) Home Scored = Average number of goals scored in Home matches
b) Home Conceded = Average number of goals conceded in Home matches
c) Away Scored = Average number of goals scored in Away matches
d) Away Conceded = Average number of goals conceded in Away matches

Next we compute the same results, but for the whole league:
e) League Home scored
f) League Home conceded
g) League Away scored
h) League Away conceded

Using above, we can assign normalized parameters to each team:
i) Home Attack strength = (a)/(e)
j) Home Defense strength = (b)/(f)
k) Away Attack strength = (c)/(g)
l) Away Defense strength = (d)/(h)

Finally, we compute the multipliers:
n) Overall goals scored Home = (e + h)/2
p) Overall goals scored Away = (f + g)/2
```

We then coded the above parameters into our python interface:

```python
# For each team - Average {HomeScored, HomeConceded, AwayScored, AwayConceded}
eplHomeTeam = eplData[['HomeTeam', 'HomeGoals', 'AwayGoals']].rename(
    columns={'HomeTeam':'Team', 'HomeGoals':'HomeScored', 'AwayGoals':'HomeConceded'}).groupby(
    ['Team'], as_index=False)[['HomeScored', 'HomeConceded']].mean()

eplAwayTeam = eplData[['AwayTeam', 'HomeGoals', 'AwayGoals']].rename(
    columns={'AwayTeam':'Team', 'HomeGoals':'AwayConceded', 'AwayGoals':'AwayScored'}).groupby(
    ['Team'], as_index=False)[['AwayScored', 'AwayConceded']].mean()

# Overall - Average {leagueHomeScored, leagueHomeConceded, leagueAwayScored, leagueAwayConceded}
leagueHomeScored, leagueHomeConceded = eplHomeTeam['HomeScored'].mean(), eplHomeTeam['HomeConceded'].mean()
leagueAwayScored, leagueAwayConceded = eplAwayTeam['AwayScored'].mean(), eplAwayTeam['AwayConceded'].mean()

eplTeamStrength = pd.merge(eplHomeTeam, eplAwayTeam, on='Team')

assert(leagueHomeScored != 0)
assert(leagueHomeConceded != 0)
assert(leagueAwayScored != 0)
assert(leagueAwayConceded != 0)

# Normalize the parameters
# For each team - {HomeAttack, HomeDefence, AwayAttack, AwayDefense}
eplTeamStrength['HomeScored'] /= leagueHomeScored
eplTeamStrength['HomeConceded'] /= leagueHomeConceded
eplTeamStrength['AwayScored'] /= leagueAwayScored
eplTeamStrength['AwayConceded'] /= leagueAwayConceded

eplTeamStrength.columns=['Team','HomeAttack','HomeDefense','AwayAttack','AwayDefense']
eplTeamStrength.set_index('Team', inplace=True)

# Overall - {overallHomeScored, overallAwayScored}
overallHomeScored = (leagueHomeScored+leagueAwayConceded)/2
overallAwayScored = (leagueHomeConceded+leagueAwayScored)/2
```

As stated before, a match between two teams can have three possible outcomes: Home team wins(H), Away team wins(A) or a Tie (T). Let the number of goals scored by the home team be 'X' and the number of goals scored by the away team be 'Y'. Then:

```
Home team wins if (X > Y)
Away team wins if (X < Y)
It's a tie if (X = Y)
```

We then proceed to calculate the probability that the match ends with the score line X-Y. Also, we decided to put a practical upper limit to the number of goals scored by a team at 10. Finally, since all the different score lines possible (for example: 4-2, 1-5, etc) are independent of each other, we can simply add up the probabilities.

```
Probability (H wins) = Σ P(X-Y scoreline) , X > Y
Probability (A wins) = Σ P(X-Y scoreline) , X < Y
Probability (Tie)    = Σ P(X-Y scoreline) , X = Y
```

Finally, we simulated a match between all the Home(H) and Away(A) teams and predicted the points scored by each team:

```python
# Predict outcome of match and assign points to the teams

def predictMatchScore(home, away):
    if home in eplTeamStrength.index and away in eplTeamStrength.index:
        lambdH = eplTeamStrength.at[home,'HomeAttack'] * eplTeamStrength.at[away,'AwayDefense'] * overallHomeScored
        lambdA = eplTeamStrength.at[away,'AwayAttack'] * eplTeamStrength.at[home,'HomeDefense'] * overallAwayScored
        probH, probA, probT = 0, 0, 0  # Probability of Home win(H), Away win(A) or Tie(T)
        for X in range(0,11):
            for Y in range(0, 11):
                p = poisson.pmf(X, lambdH) * poisson.pmf(Y, lambdA)
                if X == Y:
                    probT += p
                elif X > Y:
                    probH += p
                else:
                    probA += p
        scoreH = 3 * probH + probT
        scoreA = 3 * probA + probT
        return (scoreH, scoreA)
    else:
        return (0, 0)
```

## Putting it all together

To come up with the final standings then, we simply simulated all the league matches using the model and added up the predicted point scores to the build the points table.

```python
# Simulate the matches to predict final standings
for index, row in eplMatchesLeft.iterrows():
    home, away = row['HomeTeam'], row['AwayTeam']
    assert(home in eplPointsTable.Team.values and away in eplPointsTable.Team.values)
    sH, sA = predictMatchScore(home, away)
    eplPointsTable.loc[eplPointsTable.Team == home, 'Points'] += sH
    eplPointsTable.loc[eplPointsTable.Team == away, 'Points'] += sA
```

## Manchester City statistics

*Before their takeover by Sheikh Mansour in the 2008/09 season:*

- League position (2008/09): 10
- Average statistics from the 2005/06 season to 2008/09 season:

```
eplTeamStrength.iloc[[11]]
```

| Team | HomeAttack | HomeDefense | AwayAttack | AwayDefense |
|---|---|---|---|---|
| Man City | 0.797301 | 0.890951 | 0.958914 | 1.005413 |

*The year following their takeover by Sheikh Mansour in the 2008/09 season:*

- League position (2009/10): 5
- Average statistics from the 2009/10 season:

```
eplTeamStrength.iloc[[11]]
```

| Team | HomeAttack | HomeDefense | AwayAttack | AwayDefense |
|---|---|---|---|---|
| Man City | 1.271318 | 0.980392 | 1.568627 | 0.775194 |

**Newcastle United statistics:**

*Before their takeover by the Saudi group in the 2019/20 season:*

- League position: 13
- Average statistics from the 2015/16 season to 2019/20:

```
eplTeamStrength.iloc[[17]]

              HomeAttack   HomeDefense   AwayAttack   AwayDefense
    Team
Newcastle      0.945096      0.890281     0.791403      1.006886
```

- Predicted ranking on the basis of individual match results for the upcoming 2020/2021 Premier League season:

```
'PREDICTED FINAL STANDINGS'

              Team      Points
1          Man City   86.176216
2         Tottenham   79.079876
3         Liverpool   78.178282
4           Chelsea   70.190228
5           Arsenal   70.135188
6        Man United   68.836431
7         Leicester   56.241918
8           Everton   55.536409
9            Wolves   53.311891
10         West Ham   49.405495
11      Southampton   48.154217
12   Crystal Palace   47.090594
13         Newcastle  45.066317
14       Bournemouth  43.825595
15          Burnley   43.431080
16          Watford   42.722739
17         Brighton   37.082726
18           Cardiff  30.480337
19           Fulham   26.180560
20      Huddersfield  25.297979
```

After comparing the average scoring statistics of Manchester City and Newcastle United for four years before their takeover, we observed that their performance was nearly the same. Based on this result, we decided to assume that in the year following their takeover, their player transfers would be almost similar to Manchester City's after their budget increased. Newcastle United will focus on improving their attack by bringing in high-profile forwards and attacking midfielders. Based on this deduction, we decided to use the statistics from City's 2009/10 season to predict the performance of Newcastle United post their takeover on the basis of their similar past ranking and performance. The following results were obtained:

```
'PREDICTED FINAL STANDINGS'
           Team      Points
1       Man City   85.485361
2      Tottenham   78.336459
3      Liverpool   77.413164
4        Chelsea   69.288121
5        Arsenal   69.280216
6     Man United   68.031986
7       Newcastle  63.786796
8      Leicester   55.235902
9        Everton   54.509422
10        Wolves   52.335974
11      West Ham   48.330840
12    Southampton  47.135310
13  Crystal Palace 46.091163
14    Bournemouth  42.716583
15       Burnley   42.459014
16       Watford   41.685581
17      Brighton   36.108733
18        Cardiff  29.515840
19        Fulham   25.204467
20    Huddersfield 24.492315
```

Post their takeover, Newcastle are slated to move from the 13[th] position to the 7[th] position. This change is quite similar to Manchester City's move from the 10[th] to the 5[th] position after their takeover.

The most critical advantage of using the Poisson distribution for predicting match results is that we get to take into account the current form and statistics of other teams. In case we try to apply linear regression to predict a team's PL ranking based on their transfer budget, we fail to take into account the results of individual matches and have to use a relatively small sample size too.

**What all factors made the Saudi PIF drop out of the Newcastle deal:**

A variety of problems had arisen since April when Newcastle's Saudi takeover seemed to gain momentum.

Pressure from broadcasters and human rights organisations as well as other Premier League clubs had erected hurdles in the negotiations, it was reported. These obstacles caused delays and dissatisfaction, with the consortium mentioning that "time itself became an enemy of the transaction" in its declaration.

**1. Piracy**

A main problem in the controversy between beIN Sports and the TV charges for broadcasting piracy. Qatar 's business was the Premier League business.

In June, the BBC reported that Angus MacNeil, a member of the British Parliament, had sent to the government a letter condemning Saudi Arabia for its supposed pirates and calling on them to postpone the takingover.

Although the Saudi authorities seemed to be grappling with the issue, beIN Sports was barred from operating in Saudi Arabia in July and an arrangement has reportedly been brought closer.

In a statement beIN said: "We will wonder how Saudi people in this 'proliferative' ban on permitted broadcasting in the Prime Minister league can legally see Premier League matches in Saudi Arabia just like us for three years."

The rivalry between Qatari and the government of Saudi Arabia is part of a continuing dispute which has taken place for several long years now, some of which was marked by a "cold war."

Staveley denied that when the transaction was signed, the acquisition issue persisted. "The hacking was not a problem, but we have still been working to fix it," she said in The Times.

## 2. Opposition from Premier League Clubs

Staveley also suggested that the appropriate bid was thwarted by a number of Premier League clubs who she claimed "didn't want it to happen."

The Times notes that they had Tottenham and Liverpool. The clubs didn't really know what they faced.

## 3. Amnesty International and Human Rights Violations

Amnesty International claims that, due to the numerous financial and legal challenges involved with the coverage conflict, Premier League faced being "patsy" if it didn't avert the take-over.

In a letter to CEO Richard Masters of Premier League, Kate Allen, United Kingdom 's director of Amnesty, said to find out if NUFC owners and managers are compliant with requirements which will preserve the reputation and prestige of the game.

If the Crown Prince is the NUFC's beneficial owner by his control over Saudi Arabia's economic relations and the influence over Saudi Arabia 's sovereign fund, how can this improve the reputation and the picture of the Premier Ligue? Issues such as these forced officials of the Premier League to get out of the deal.

As long as these issues were not addressed, many assumed that, by working with the nation whose actions defied international law, and by using the glamor and glory of football as a tool for deep-rooting moralistic acts, the Premier League would be in danger of moving against the ideas of the global football world.

In June the BBC reported that Hatiz Cengiz, the fiancee of the murdered journalist Jamal Khashoggi, had also raised issues with Premier League leaders.

## 4. Uncertainty caused by COVID 19

In addition to all these diplomatic considerations, economic conditions have also played an crucial part in the cancelation of the contract. Owing to COVID 19, the instability persists in all markets and soccer markets are no exception, UEFA 2020 has been delayed, the first league has taken more than two months, and thus market turmoil has prevented Saudi citizens from cancelling transactions.

In addition, oil prices were not stable, because the oil market, Saudi Arabian's biggest cash cow, had been hit by a decline in the demand for oil due to a COVID crisis, which prompted many countries to impose tight lockouts, resulting in a major decrease in aggregate demand. It can therefore be inferred that Saudi, compelled to withdraw from the swap agreement by political reasons coupled to the economic reasons induced by COVID.

## Some Limitations & Scope for improvement

The model is based on Poisson's distribution; hence it inherits some limitation and constraints under which it has to work:

(a) Events are independent of each other. So, if a Team A scores a goal, the probability of scoring the second goal won't be affected by the previous goal. In reality the case is completely different, after scoring or conceding a goal, the strategy of both the team involved changes.

(b) The matches are independent, and all matches are equally important for the teams, in real world situation, performance in the previous match have huge impact on the strategy and psyche of the players.

(c) Manager plays a vital role in the performance of a club, they are the coach and strategist for the team, hence the "Performance of a manager" is an important variable that should be considered in a model.

Our future work would be on improving our model and making it more accurate with respect to the "Real World" situation. Furthermore, we are also working on a model using "difference in difference" approach to analyse the impact of Manchester City's ownership transfer.

The benefits of using difference in difference approach would be:

(i) The method is fairly intuitive as well as flexible.

(ii) If basic assumptions are met, it can be useful to establish a "causal relationship".

(iii) Variables like "Performance of the manager" would be easier to introduce.

(iv) The method accounts for the for change due to factors other than the treatment or intervention being studied. Hence it would provide more accurate picture of the Real football world.